\documentclass[twocolumn,showpacs,prl]{revtex4}
\usepackage{graphicx}% Include figure files

\begin{document}

\preprint{}
\title{
Vortex state in double transition superconductors
}

\author{Yasushi Matsunaga} 
\email{matsunaga@mp.okayama-u.ac.jp}
\author{Masanori Ichioka} 
\email{oka@mp.okayama-u.ac.jp}
\author{Kazushige Machida}
\email{machida@mp.okayama-u.ac.jp}
\affiliation{Department of Physics, Okayama University,
         Okayama 700-8530, Japan}

\date{
\today
%26 October, 2003
}

\begin{abstract}
Novel vortex phase and nature of double transition field are investigated 
by two-component Ginzburg-Landau theory in a situation where  
fourfold-twofold symmetric superconducting double transition occurs. 
The deformation from $60^\circ$ triangular vortex lattice 
and a possibility of the vortex sheet structure are discussed. 
In the presence of the gradient coupling, the transition 
changes to a crossover at finite field. 
These characters are important to identify the multiple superconducting 
phase in ${\rm PrOs_4Sb_{12}}$. 
\end{abstract}

\pacs{74.25.Qt, 74.20.Rp, 74.20.De, 74.25.Dw} 

% 74.25.Op Mixed states, critical fields, and surface sheaths
% 74.25.Qt Vortex lattices, flux pinning, flux creep
% 74.25.Dw Superconductivity phase diagrams
% 74.20.De Phenomenological theories (two-fluid, Ginzburg-Landau, etc.)
% 74.20.Rp Pairing symmetries (other than s-wave)
% 74.70.Tx Heavy-fermion superconductors

\maketitle 
%%%%%%%%%%%%%%%%%%%%%%%%%%%%%%%%%%%%%%%%%%%%%%%%%%%%%%%
%\section{Introduction}
%\label{sec:introduction}

%\section{formulation} 

In the newly discovered superconductor ${\rm PrOs_4Sb_{12}}$, 
which is a heavy-fermion compound with filled skutterudite 
structure, a novel pairing mechanism is considered 
in connection to quadrupole fluctuations~\cite{Bauer}.  
Several experimental results suggest unconventional pairing symmetry. 
In the nuclear spin-lattice relaxation experiment, 
the coherence peak is absent below the transition temperature $T_{\rm c}$, 
and the low temperature behavior indicates 
that ${\rm PrOs_4Sb_{12}}$ has full gap or point nodes, 
excluding line nodes, at a zero field~\cite{Kotegawa}.  
The specific heat jump at  $T_{\rm c}$  
shows double transition~\cite{Vollmer}.  
The thermal transport measurement in the magnetic field rotated 
in the $ab$-plane of the crystal axes reveals the phase diagram of the 
double transition,
and shows that a fourfold-symmetric pairing function 
around the $c$-axis in the high-field H-phase is changed 
to a twofold-symmetric one in the low-field L-phase 
at the second transition field $H^\ast$, 
when magnetic field or temperature is decreased~\cite{Izawa}. 
Thus, the $H$-$T$ phase diagram is divided into two regions,  
H- and L-phases. 
The muon spin relaxation ($\mu$SR) experiment reports the spontaneous moment 
in the superconducting phase~\cite{Aoki}.  
These results indicate an unconventional pairing of the 
superconductivity in ${\rm PrOs_4Sb_{12}}$. 

To explain the double transition, we need multiple components  
for the pairing functions.   
The fourfold-twofold transition in ${\rm PrOs_4Sb_{12}}$ 
can be explained as follows: 
After the fourfold-symmetric pairing component appears at the first transition,
the second component appears at the second transition, 
and the combination of these components gives rise to 
the twofold-symmetric gap structure. 
For ${\rm PrOs_4Sb_{12}}$, 
the scenario of ``anisotropic-$s$''+${\rm i}d$-wave pairing 
with point nodes was proposed, and the double transition is analyzed 
by the Ginzburg-Landau (GL) theory at a zero field~\cite{Goryo}.    
In the vortex states at finite fields, however, 
this fourfold-twofold transition is not a true transition, 
when the gradient coupling terms are present in the GL theory~\cite{Ichioka}.  
That is, the twofold component at the L-phase is induced by the 
gradient coupling, and produces twofold symmetric behavior 
in addition to fourfold symmetric one, even in the H-phase above $H^\ast$. 
To avoid the mixing of the twofold component above $H^\ast$, 
the plausible pairing symmetry should be a triplet pairing 
where $d$-vector of the two components are orthogonal each other 
(i.e., ${\bf d}^\ast_1\cdot {\bf d}_2 =0$), so that 
the gradient coupling terms vanish~\cite{Ichioka}. 

On the other hand, in multi-component superconductors, 
it is interesting to examine the possibility of exotic vortex states, 
such as a coreless 
vortex~\cite{Tokuyasu,Machida,Hirano,Fujita,Kita,Salomaa,Parts}. 
In this letter, we investigate the double transition by the 
two-component GL theory in the magnetic field, 
and  calculate the vortex structure 
by the time-evolution of the time-dependent GL (TDGL) equation. 

First, we derive two-component GL equation appropriate to 
study the fourfold-twofold transition in 
${\rm PrOs_4Sb_{12}}$. 
The pair potential in $2 \times 2$ matrix form is decomposed 
to two components as  
$\hat\Delta({\bf r},{\bf k})
=\eta_1({\bf r})\hat\phi_1({\bf k})
+\eta_2({\bf r})\hat\phi_2({\bf k})$  
with the order parameter $\eta_m({\bf r})$, 
where ${\bf r}$ is the center of mass coordinate of the Cooper pair 
and $m=1,2$. 
The relative momentum ${\bf k}$ of the pair is mapped on the Fermi surface. 
The pairing function is given by 
$\hat\phi_m({\bf k})={\rm i}\hat\sigma_y\phi_m({\bf k})$ 
for the singlet pairing, and 
$\hat\phi_m({\bf k})={\rm i}\sum_{j=x,y,z}
d_{m,j}({\bf k})\hat\sigma_j \hat\sigma_y $ 
for the triplet pairing 
with Pauli matrices $\hat{\sigma}_x$, $\hat{\sigma}_y$, $\hat{\sigma}_z$.  
We assume that the superconducting gap by $\hat\phi_1({\bf k})$ 
($\hat\phi_2({\bf k})$) has fourfold 
(twofold) symmetry~\cite{Goryo,Ichioka},  
and that the transition temperature estimated from 
the pairing interaction is lower for the second component, 
i.e., $T_{\rm c}=T_{{\rm c}1}>T_{{\rm c}2}$. 
Since the pairing symmetry for ${\rm PrOs_4Sb_{12}}$ is not 
established yet, the pairing function forms $\hat\phi_m({\bf k})$ 
are not specified in this study. 

Within the GL approximation, 
the free energy in the superconducting state is generally given by 
$F_s=F_n+\int f({\bf r}){\rm d}{\bf r}$ 
with 
\begin{eqnarray} && 
f({\bf r})=-\alpha_0(T_{{\rm c}}-T)|\eta_1|^2 
           -\alpha_0(T_{{\rm c}2}-T)|\eta_2|^2 
\nonumber \\ && 
+|A_2|\left\{ 
 \langle \frac{1}{2}{\rm tr}(\hat\Delta^\dagger  
   ({\bf v}\cdot{\bf q})^2 \hat\Delta)\rangle 
+\langle \frac{1}{2}
{\rm tr}(\hat\Delta^\dagger \hat\Delta
         \hat\Delta^\dagger \hat\Delta)\rangle \right\} \qquad 
\label{eq:fe1}
\end{eqnarray} 
in the clean limit~\cite{Schopohl}, 
where ${\bf q}=(\hbar/{\rm i})\nabla+(2\pi/\phi_0){\bf A}$, 
$F_n$ is the free energy in the normal state, 
$|A_2|=7\zeta(3)/(16 \pi^2 T_{{\rm c}}^2)$ with Riemann's $\zeta$-function, 
$\phi_0$ is a flux quantum, ${\bf v}$ is a Fermi velocity, 
$\langle\cdots\rangle$ indicates the Fermi surface average of ${\bf k}$, 
and ${\bf A}$ is a vector potential. 
Since the magnetic field is applied along the $z$-axis, $q_z=0$. 

In the dimensionless form, Eq. (\ref{eq:fe1}) is written as 
\begin{eqnarray} && 
\tilde{f} \equiv \frac{f}{f_0}
=-\left(1-\frac{T}{T_{{\rm c}}}\right)|\eta_1|^2 
 -\left(\frac{T_{{\rm c}2}}{T_{{\rm c}}}
       -\frac{T}{T_{{\rm c}}}\right)|\eta_2|^2 
\nonumber \\ && 
+ \eta_1^\ast (q_x^2+q_y^2) \eta_1 
+C_{22x} \eta_2^\ast q_x^2 \eta_2 
+C_{22y} \eta_2^\ast q_y^2 \eta_2 
\nonumber \\ && \hspace{0.5cm} 
+C_{12x} \eta_1^\ast q_x^2 \eta_2 +C_{12x}^\ast \eta_2^\ast q_x^2 \eta_1  
\nonumber \\ && \hspace{0.5cm}  
+C_{12y} \eta_1^\ast q_y^2 \eta_2 +C_{12y}^\ast \eta_2^\ast q_y^2 \eta_1  
\nonumber \\ && 
+\frac{1}{2} \{ |\eta_1|^4 +C_2 |\eta_2|^4 +4C_3
 |\eta_1|^2  |\eta_2|^2  
\nonumber \\ && \hspace{0.5cm}  
+C_4^\ast \eta_1^{\ast 2} \eta_2^2 
+C_4      \eta_2^{\ast 2} \eta_1^2 \},  \qquad 
\label{eq:fe2}
\end{eqnarray} 
using 
$f_0=\alpha_0 T_{{\rm c}} \eta_0^2 $, 
$\eta_0^2=\alpha_0 T_{{\rm c}}/(2|A_2| C_1) \equiv 1$ and  
$\xi_0^2=|A_2| C_{11x}/(\alpha_0 T_{{\rm c}}) \equiv 1$. 
The coefficients of the gradient terms in Eq. (\ref{eq:fe2}) are given by 
\begin{eqnarray} && 
C_{11x}=\langle \overline{|\phi_1|^2} v_x^2\rangle 
       =\langle \overline{|\phi_1|^2} v_y^2\rangle , 
\nonumber \\ && 
C_{22x}= \langle \overline{|\phi_2|^2} v_x^2 \rangle /C_{11x}
\equiv (1-c)/(X\sqrt{1-c^2}) , 
\nonumber \\ && 
C_{22y}= \langle \overline{|\phi_2|^2} v_y^2 \rangle /C_{11x}
\equiv (1+c)/(X\sqrt{1-c^2}) , 
\nonumber \\ && 
C_{12x}= \langle \overline{\phi_1^\ast \phi_2} v_x^2 \rangle /C_{11x} , 
\quad 
C_{12y}= \langle \overline{\phi_1^\ast \phi_2} v_y^2 \rangle /C_{11x} , 
\qquad
\label{eq:grd}
\end{eqnarray}  
where 
$\overline{\phi_m^\ast \phi_n}\equiv\phi_m^\ast \phi_n$ 
for the singlet pairing, 
$\overline{\phi_m^\ast \phi_n}\equiv{\bf d}_m^\ast\cdot{\bf d}_n$ 
for the triplet pairing and 
$\overline{|\phi_m|^2} \equiv \overline{\phi_m^\ast \phi_m}$. 
The slope ratio of $H^\ast$ and $H_{{\rm c}2}$ in the $H$-$T$ phase 
diagram is roughly given by $X=[ 
\langle \overline{|\phi_1|^2} v_x^2 \rangle
\langle \overline{|\phi_1|^2} v_y^2 \rangle / 
\langle \overline{|\phi_2|^2} v_x^2 \rangle
\langle \overline{|\phi_2|^2} v_y^2 \rangle
]^{1/2} $. 
To reproduce the experimentally obtained $H^\ast$ behavior, 
it is appropriate to use $X \sim H^\ast/H_{{\rm c}2}\sim  0.5$. 
In Eq. (\ref{eq:grd}), $c$ is an anisotropic parameter related to 
the second order parameter. 
When $\overline{|\phi_2|^2}$ has two-fold symmetry under the rotation 
around the $z$-axis, $c$ is not zero 
since $\langle \overline{|\phi_2|^2} v_x^2 \rangle
\ne \langle \overline{|\phi_2|^2} v_y^2 \rangle $. 
$C_{12x}$ and $C_{12y}$ reflect the strength of the gradient coupling  
between $\eta_1$ and $\eta_2$.  

The coefficients of the quadratic terms in Eq. (\ref{eq:fe2}) are given by 
$C_1=\langle |\phi_1|^4 \rangle$, 
$C_2=\langle |\phi_2|^4 \rangle/C_1$, 
$C_3=\langle |\phi_1|^2|\phi_2|^2 \rangle/C_1$,  
$C_4=\langle \phi_1^{\ast 2} \phi_2^2 \rangle/C_1$ 
for the singlet pairing, and 
$C_1=\langle 2|{\bf d}_1|^4 -|{\bf d}_1 \cdot{\bf d}_1|^2 \rangle$, 
$C_2=\langle 2|{\bf d}_2|^4 -|{\bf d}_2 \cdot{\bf d}_2|^2 \rangle/C_1$, 
$C_3=\langle |{\bf d}_1|^2|{\bf d}_2|^2+|{\bf d}_1^\ast\cdot{\bf d}_2|^2 
-|{\bf d}_1\cdot{\bf d}_2|^2 \rangle/C_1$, 
$C_4=\langle 2({\bf d}^\ast_2 \cdot {\bf d}_1)^2 
    -({\bf d}^\ast_2 \cdot {\bf d}^\ast_2)
     ({\bf d}_1 \cdot {\bf d}_1)\rangle/C_1 $
for the triplet pairing. 
However, we can not identify the definitive values of 
coefficients in Eq. (\ref{eq:fe2}) for ${\rm PrOs_4 Sb_{12}}$, 
because the detailed information of the pairing function 
and Fermi surface structure have not been established yet.     
It is noted that the Fermi surface anisotropy also  largely affects 
on the coefficients. 
Therefore, we treat these coefficients as arbitrary parameters, 
and report some typical results obtained in this framework. 

Before considering vortex states, we study a uniform state 
at a zero field. 
From the free energy minimum condition, 
the relative phase of $\eta_1$ and $\eta_2$ becomes 
$(\alpha\pm\pi)/2$ with $\alpha$ given by 
$C_4=|C_4|{\rm e}^{{\rm i}\alpha}$. 
For the H-phase, 
$\eta_1=(1-T/T_{{\rm c}})^{1/2}$ and $\eta_2=0$. 
The second component $\eta_2$ appears at the lower transition 
temperature $T^\ast$ given by  
\begin{eqnarray}
\frac{T^\ast}{T_{\rm c}}
=\frac{T_{{\rm c}2}/T_{{\rm c}}-(2C_3 -|C_4|) }{ 1 -(2C_3 -|C_4|) }, 
\end{eqnarray}
which is derived by linearizing the equation 
$\partial \tilde{f}/\partial \eta^\ast_2=0$. 
In the presence of $\eta_1$, $T^\ast$ is suppressed 
compared with $T_{{\rm c}2}$. 
To assure that $T^\ast>0$, we have to satisfy 
$2C_3 -|C_4|< T_{{\rm c}2}/T_{{\rm c}}$. 

The vortex structure is calculated by the time-evolution following 
the TDGL equation coupled with Maxwell equation~\cite{Kato,MMachida}, 
\begin{eqnarray} && 
\frac{\partial}{\partial t}\eta_1 
= -\frac{1}{12} \frac{\partial \tilde{f}}{\partial \eta_1^\ast}, 
\qquad 
\frac{\partial}{\partial t}\eta_2 
= -\frac{1}{12} \frac{\partial \tilde{f}}{\partial \eta_2^\ast}, 
\label{eq:TDGL}
\\ && 
\frac{\partial}{\partial t}{\bf A} =  \tilde{\bf j}_{\rm s} 
-\kappa^2 \nabla\times{\bf B}, 
\qquad 
{\bf B}=\nabla\times{\bf A}. 
\label{eq:Maxwell} 
\end{eqnarray} 
The supercurrent $\tilde{\bf j}_{\rm s}
=(\tilde{j}_{{\rm s},x},\tilde{j}_{{\rm s},y})
\propto(\partial \tilde{f}/\partial A_x, \partial \tilde{f}/\partial A_y)$ 
is given by 
\begin{eqnarray} &&
\tilde{j}_{{\rm s},x}
={\rm Re} [ 
         \eta_1^\ast (q_x \eta_1) 
+C_{22x} \eta_2^\ast (q_x \eta_2) 
\nonumber \\ && \hspace{1cm} 
+C_{12x} \eta_1^\ast (q_x \eta_2) 
+C_{12x}^\ast \eta_2^\ast (q_x \eta_1)  ],  \qquad 
\label{eq:jsx}
\\ && 
\tilde{j}_{{\rm s},y}
={\rm Re}[  \eta_1^\ast (q_y \eta_1 )
+C_{22y} \eta_2^\ast (q_y \eta_2 )
\nonumber \\ && \hspace{1cm} 
+C_{12y} \eta_1^\ast (q_y \eta_2) 
+C_{12y}^\ast \eta_2^\ast (q_y \eta_1)  ]. 
\label{eq:jsy}
\end{eqnarray} 
We use the same scale units as in Refs. \onlinecite{Kato} and 
\onlinecite{MMachida} for length, field and time,  
except for the order parameters. 
We here scale $\eta_m$ by $\eta_0$ instead of 
$\eta_0(T)=\eta_0(1-T/T_{{\rm c}})^{1/2}$. 
In our calculations, we typically use the GL parameter 
$\kappa=4$, which belongs to a high-$\kappa$ case, i.e., 
the order parameter structure is not significantly affected 
by the internal field distribution. 

Calculations are performed in a two-dimensional square  area. 
Outside the open boundary, 
we set $\eta_1=\eta_2=0$ and $B({\bf r})=H$ with an applied field $H$. 
For the initial state of $\eta_1$, $\eta_2$ and ${\bf A}$ inside, 
we use a uniform state at a zero-field. 
Through the time evolution, vortices penetrate from the boundary. 
After enough time later, the time-evolution converges 
and the vortex lattice state is obtained. 
We analyze this final state. 

%%%%%%%%%%%%%%%%%%%%%%
\begin{figure} [tbh]
\includegraphics[width=8.0cm]{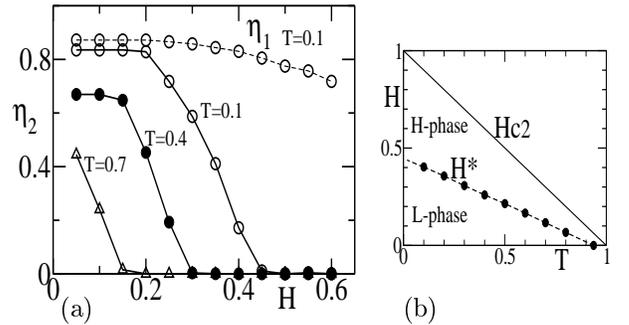} 
\caption{
(a) Maximum of $|\eta_2({\bf r})|$ (solid lines)  
as a function of $H/H_{{\rm c}2}(0)$. 
$T/T_{\rm c}=0.1(\circ)$, 0.4($\bullet$) and 0.7($\triangle$). 
$\eta_2$ appears at $H<H^\ast(T)$.  
For $T/T_{\rm c}=0.1$, $H$-dependence of $|\eta_1|$ is 
also presented (a dashed line). 
(b) $H$-$T$ phase diagram in this GL theory. 
The transition field $H^\ast$ and $H_{{\rm c}2}$ are presented.  
} 
\label{fig:1}
\end{figure} 
%%%%%%%%%%%%%%%%%%%%%%

We discuss the case when the gradient coupling 
is absent ($C_{12x}=C_{12y}=0$). 
Here, we set $T_{{\rm c}2}/T_{\rm c}=0.95$, 
$C_2=1$, $C_3=-C_4=0.2$, $c=0.3$ and $X=0.5$. 
In Fig. \ref{fig:1}(a), 
we show the maximum order parameter amplitude   
at $T/T_{{\rm c}}=0.1$, 0.4 and 0.7 
as a function of the external field $H/H_{{\rm c}2}(0)$ 
with $H_{{\rm c}2}(0)=\phi_0/(2 \pi \xi_0^2)$. 
At a critical field $H^\ast(T)$, the second component $\eta_2$ 
appears as a second order transition. 
The $T$-dependence of $H^\ast$ is shown in Fig. \ref{fig:1}(b). 
This $H$-$T$ phase diagram qualitatively reproduces the double transition 
of ${\rm PrOs_4Sb_{12}}$~\cite{Izawa}. 

%%%%%%%%%%%%%%%%%%%%%%
\begin{figure} [tbh]
\includegraphics[width=8.0cm]{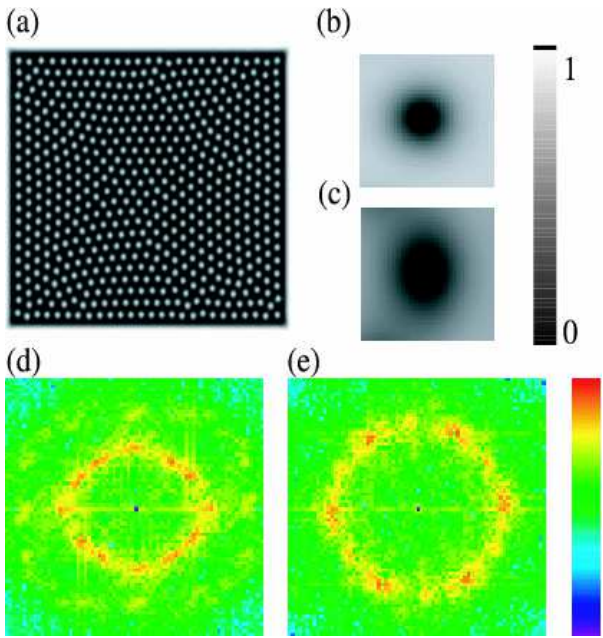} 
\caption{
(a) Density plot of an internal field distribution $B({\bf r})$ 
at $H/H_{{\rm c}2}(0)$=0.2 and $T/T_{\rm c}=0.1$. 
White region corresponds to the vortex core with large  $B({\bf r})$. 
Density plot of $|\eta_1({\bf r})|$ (b) and $|\eta_2({\bf r})|$ (c) 
around a vortex. 
Diffraction pattern $|B_{\bf q}|^2$ 
at $H/H_{{\rm c}2}(0)$=0.2 [L-phase] (d) and 0.5 [H-phase] (e). 
}
\label{fig:2}
\end{figure} 
%%%%%%%%%%%%%%%%%%%%%%

At $H/H_{{\rm c}2}(0)=0.2$ and $T/T_{{\rm c}}=0.1$, 
the vortex distribution is presented in Fig. \ref{fig:2}(a), 
where the internal field distribution is shown as a density plot.
There are some domains of the different orientation of 
the triangular lattice, since the vortex lattice configuration 
is affected by the boundary. 
And the triangular lattice is deformed from $60^\circ$ triangle 
by the effect of finite $c$ in Eq. (\ref{eq:grd}), 
coming from the twofold symmetry of $\hat{\phi}_2$. 
The vortex core shapes of $|\eta_1({\bf r})|$ and $|\eta_2({\bf r})|$ are, 
respectively, shown in Figs. \ref{fig:2}(b) and \ref{fig:2}(c). 
Vortex core shape of $|\eta_2({\bf r})|$ is stretched out toward 
the $y$-direction, 
while that of $|\eta_1({\bf r})|$ remains to be circular. 
The deformation of the vortex lattice is clear, when 
we see the form factor of the internal field distribution, 
as in the neutron scattering experiments. 
Figure \ref{fig:2}(d) shows the ``diffraction pattern'' $|B_{\bf q}|^2$ 
in the L-phase, with the Fourier component 
$B_{\bf q}=\sum_{\bf r}{\rm e}^{{\rm i}{\bf q}\cdot{\bf r}}B({\bf r})$, 
where we exclude the region near the boundary in the integration 
over ${\bf r}$.  
There appear six peaks of the triangular lattice, 
and the rotated ones for different orientations. 
These peaks are on an ellipse with long axis $2q_x$ 
and short axis $2q_y$ with the ratio $q_x/q_y \sim 1.2$. 
For a reference, we also show the diffraction pattern 
of the $60^\circ$ triangular lattice in the H-phase 
in Fig. \ref{fig:2}(e), where peaks appear on a circle.  
The observation of this vortex lattice deformation  
may be another means to detect the fourfold-twofold 
transition in the $H$-$T$ phase diagram. 

%%%%%%%%%%%%%%%%%%%%%%
\begin{figure} [tbh]
\includegraphics[width=8.5cm]{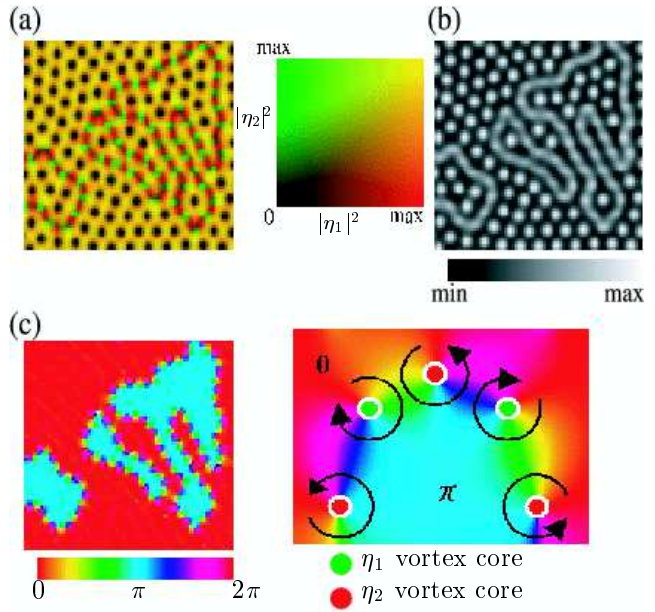} 
\caption{
Vortex structure at $H/H_{{\rm c}2}(0)=0.3$ and $T/T_{\rm c}=0.1$ [L-phase]. 
Inside region is focused. 
(a) Density plot of $|\eta_1({\bf r})|$ and $|\eta_2({\bf r})|$. 
In the black regions,  $|\eta_1|$ and $|\eta_2|$ share a vortex core. 
Green (red) regions show that 
only  $|\eta_1|$ ($|\eta_2|$) has a vortex core. 
(b) Internal field distribution $B({\bf r})$. 
(c) Relative phase ${\rm arg}\{ \eta_1({\bf r})/\eta_2({\bf r}) \}$. 
Right panel schematically shows relative phase near the vortex sheet 
appearing at the domain wall. } 
\label{fig:3}
\end{figure} 
%%%%%%%%%%%%%%%%%%%%%%

At higher field in the L-phase, we also find the vortex sheet 
structure~\cite{Parts} in addition to the regular vortices. 
We show the spatial distribution of 
$|\eta_1({\bf r})|$ and $|\eta_2({\bf r})|$ 
at $H/H_{{\rm c}2}(0)=0.3$ in Fig. \ref{fig:3}(a). 
The black circle region presents regular vortex, 
where $|\eta_1({\bf r})|$ and $|\eta_2({\bf r})|$ share the same vortex core, 
as seen in Fig. \ref{fig:2}(a). 
The green (red) circle region shows the $\eta_1$- ($\eta_2$-) vortex, 
where only $|\eta_1({\bf r})|$ ($|\eta_2({\bf r})|$) has vortex core 
and the other $|\eta_2({\bf r})|$ ($|\eta_1({\bf r})|$) does not. 
These green and red vortex cores are located alternatively along a loop, 
forming vortex sheet. 
In the internal field distribution shown in Fig. \ref{fig:3}(b), 
$B({\bf r})$ has a sharp localized peak at a regular vortex. 
And large $B({\bf r})$ regions of the vortex core on the vortex sheet 
are connected each other along the vortex sheet. 
Each of $\eta_1$- and $\eta_2$-vortices has half flux quantum. 
If this line structure is found by the direct observation of the 
internal field distribution, 
it can be evidence of the vortex sheet appearing in unconventional 
superconductors. 
The relative phase of  $\eta_1({\bf r})$ and $\eta_2({\bf r})$ are presented 
in Fig. \ref{fig:3}(c). 
Around the regular vortex, the relative phase is fixed at 
$(\alpha+\pi)/2$ or $(\alpha-\pi)/2$ (in our parameter, $\alpha=\pi$). 
Across the vortex sheet, the relative phase changes from $0$ (red region) to 
$\pi$ (blue region). 
This indicates that the vortex sheet appears at the domain wall 
between the region with the relative phase $(\alpha+\pi)/2$ and 
that with  $(\alpha-\pi)/2$, 
and that it is not easy to disconnect vortices on the vortex sheet. 
As shown in the right panel in Fig. \ref{fig:3}(c), 
since windings of the relative phase are opposite at the $\eta_1$-vortex 
and the $\eta_2$-vortex,  
the relative phase changes from $0$ to $\pi$ through $\pi/2$ (yellow) or 
$3\pi/2$ (purple) alternatively between the nearest neighbor vortices 
along the vortex sheet. 
Since domains with relative phase $(\alpha\pm\pi)/2$ are degenerate 
in free energy, we can expect both domains to coexist in sample materials. 
In the presence of the domain wall between domains, vortex sheet 
appears when applying field. 
In our simulation for the penetration process of vortices, 
boundary region helps creation of the domain wall 
with vortex sheet structure. 
The domain wall and vortex sheet are 
tight structure when coming inside, and survive stably.

%%%%%%%%%%%%%%%%%%%%%%
\begin{figure} [tbh]
\includegraphics[width=4.5cm,height=3.5cm]{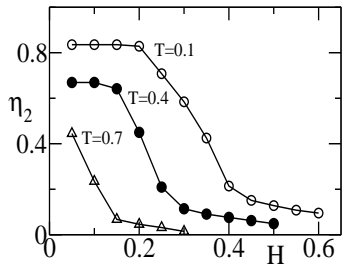} 
\caption{
Maximum of $|\eta_2({\bf r})|$ 
as a function of $H/H_{{\rm c}2}(0)$
in the presence of the gradient coupling. 
$T/T_{\rm c}=0.1(\circ)$, 0.4($\bullet$) and 0.7($\triangle$).   
$\eta_2$ survives up to $H_{{\rm c}2}(T)$. 
} 
\label{fig:4}
\end{figure} 
%%%%%%%%%%%%%%%%%%%%%%

Lastly, we consider the case when the gradient coupling terms 
do not vanish. 
We show the results when $C_{12x}=0.2$ and $C_{12y}=0$ 
and the other parameters are kept same. 
We note that, if we consider the $s+{\rm i}d$-state 
(e.g. $\phi_1({\bf k})\propto 1- (k_x^4+k_y^4+k_z^4)$ and 
$\phi_2({\bf k})\propto k_x^4 -k_z^4$)~\cite{Goryo}, $C_{12y}=0$. 
We see the similar vortex structure 
as in Figs. \ref{fig:2} and \ref{fig:3} in the L-phase. 
The field dependence of the order parameter maximum 
of $|\eta_2({\bf r})|$ is shown in Fig. \ref{fig:4}. 
On raising field, $|\eta_2|$ decreases, but it does not vanish at $H^\ast$. 
$|\eta_2|$ has finite value up to $H_{{\rm c}2}(T)$. 
Therefore, twofold symmetric order parameter 
$\eta_2({\bf r})\hat\phi_2({\bf k})$ is mixed in addition to 
the fourfold one $\eta_1({\bf r})\hat\phi_1({\bf k})$ 
in the H-phase~\cite{Ichioka}. 
That is, $H^\ast$ is not a phase transition but the crossover field 
where $\eta_2$ is enhanced in the vortex state. 
While the anomaly of the second order transition appears at $T^\ast$ 
in the zero field case, the anomaly of the phase transition is not observed 
around $H^\ast$ at finite fields.

In summary, we investigate the vortex state 
based on the two-component GL theory 
in the situation when the fourfold-twofold symmetric superconducting 
transition occurs. 
We estimate the transition field $H^\ast$ 
where second order parameter $\eta_2$ appears. 
However, $H^\ast$ is changed to a crossover field 
when the gradient coupling terms exist, 
because the small $\eta_2$ survives up to $H_{{\rm c}2}(T)$. 
In the L-phase below $H^\ast$, the vortex lattice deforms 
from $60^\circ$ triangular lattice due to the effect of 
the twofold symmetric second order parameter. 
In the two-component superconductor, 
there is a possibility to observe the exotic vortex structure 
such as vortex sheet at the domain wall, where two order parameters have 
different vortex cores and these cores are alternatively 
located along a line. 
These characters of the vortex structure may be clear evidence 
of the fourfold-twofold symmetric double transition and 
unconventional multi-component superconductivity, 
if they are experimentally observed. 

We would like to thank K. Izawa, Y. Matsuda, Y. Aoki, 
N. Nakai and M. Takigawa for fruitful discussions.

%%%% references %%%%%%%%%%%%%%%%%%%%%%%%%%%%%%%%%%%%%%%%%%%%%%%%%%%%
%\newpage

%%%%%%%%%%%%%%%%%%%%%%%%%%%%%%%%%%%%%%%%%%%%%%%%%%%%%%%%
\end{document}